\documentclass[12pt,a4paper]{article}
\usepackage{amsmath,amssymb,amscd}
\usepackage{graphicx,epsfig}

\begin{document}

\begin{center}
{\bf COLOR DECOHERENCE IN IN-MEDIUM QCD CASCADES }
\end{center}

\bigskip

\begin{center}

{\bf A.~Leonidov$^{(a,b)\,}$, V.~Nechitailo$^{(a,b)\,}$}

\bigskip

(a) {\it Theoretical Physics Department, P.N.~Lebedev Physics Institute, \\
Moscow, Russia}

(b) {\it Institute of Theoretical and Experimental Physics, Moscow, Russia}

\end{center}

\bigskip

\begin{center}
{\bf Abstract}
\end{center}

The talk, based on \cite{LN10}, analyzes the consequences of the assumption that the effects of quantum coherence and the resulting angular
ordering in QCD cascades are disrupted within the hot fireball created in ultrarelativistic heavy ion collisions.

\bigskip

Studying modifications of intrajet properties of high energy jets in nuclear collisions has drawn considerable theoretical
\cite{PYQUEN,JEWEL,QPYTHIA,QHERWIG} and experimental \cite{JSTAR,JLHC} attention. All of the listed MC generators take into account the
medium-induced radiative energy loss that is believed to be a major source of jet quenching. In addition, JEWEL \cite{JEWEL} also takes into
account the effects of rescattering of cascade particles on those in the medium assuming, in particular, that such a scattering destroys angular
ordering among the corresponding gluon emission angles.

The focus of the present work is on analyzing the consequences of the assumption that due to the violent color environment created in heavy ion
collisions the interference effects requiring precise color matching between the different contributing diagrams, in particular the one leading to
the angular ordering, get disrupted within the hot fireball having some finite size. The assumption in question is supported by the fact \cite{SG}
that rotation of the color spin of the hard mode is the fastest process in quark-gluon plasma. In particular, it is faster than the rate of the
change of the energy-momentum of the hard mode. The resulting decoherence has a significant effect on the spatiotemporal pattern of the part of
cascade that evolves inside the hot fireball and on the properties of the final cascade-generated configuration. A detailed comparison of globally
angular ordered and non-angular ordered cases was made in \cite{BS87}. In distinction to \cite{JEWEL} we do not take into account the
energy-momentum exchange with the medium accompanying randomization of color leading to decoherence and in distinction to \cite{BS87,JEWEL}
explicitly take into account the finite size of the hot zone.

\begin{figure}[h]
 \centering
 \includegraphics[width=0.75\textwidth]{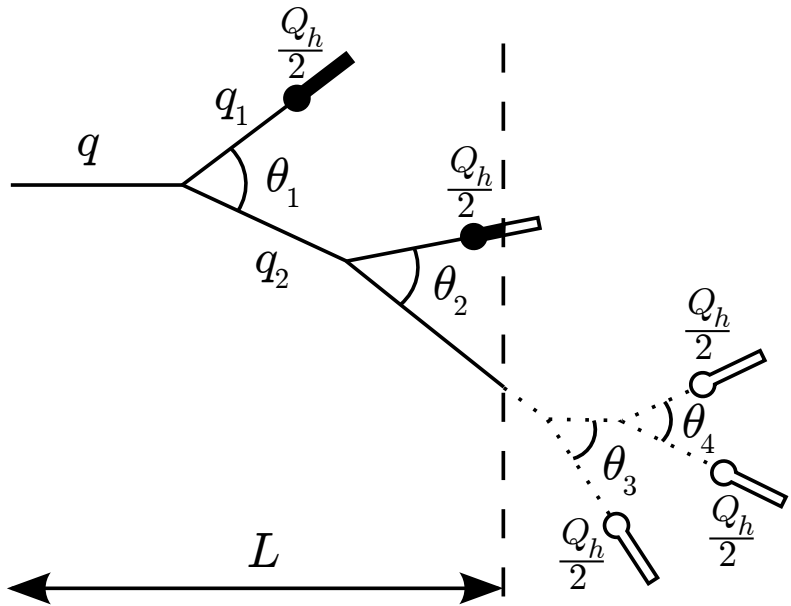}
 \caption{A sketch of the in-medium QCD cascade. The angles $\theta_3$ and $\theta_4$ of in-vacuum decays are ordered,
 $\theta_4 < \theta_3$, while those of in-medium
 decays $\theta_1$ and $\theta_2$ are not.}
 \label{fcascade}
\end{figure}

The key elements of the model used in the present work are
illustrated in Fig.~\ref{fcascade}.

To determine whether a given line is still inside the medium or
has already left it, one needs to set up a clock counting time
along the cascade. The only possibility of associating a
spatiotemporal pattern with the cascade is to use the lifetime
$\tau$ of a virtual parton which, for a parton with the energy $E$
and virtuality $Q^2$ that has been created in the decay of its
parent parton with the virtuality $Q^2_{\rm par}$, reads
\begin{equation}\label{formtime}
\tau = E \left(\frac{1}{Q^2}-\frac{1}{Q^2_{\rm par}} \right)
\end{equation}
The lifetime of the initial parton is taken to be $\tau_{\rm
in}=E_{\rm in}/Q^2_{\rm in}$. For decays taking place inside the
medium such as those of gluons with the momenta $q$ and $q_1$ the
quantum coherence effects resulting in the angular ordering are
assumed to be destroyed. The effect is taken into account by
switching off angular ordering for the decay vertices inside the
hot fireball.

\begin{figure}[h]
 \centering
 \includegraphics[width=\textwidth]{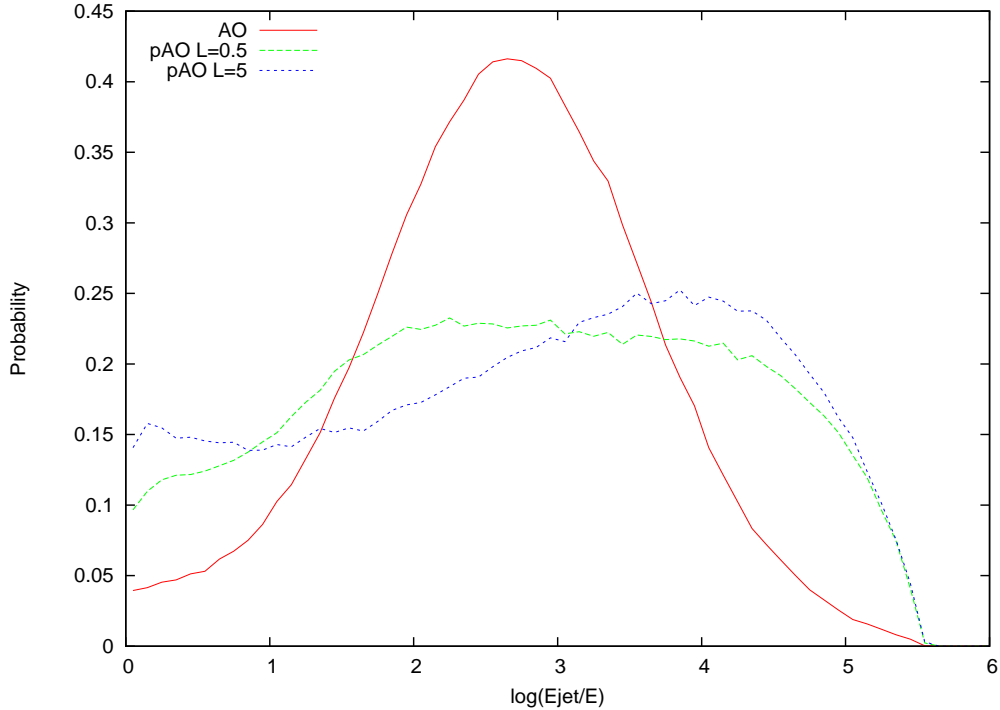}
 \caption{Rapidity distribution $P(y)$ of final prehadrons:
1) $L=0\,{\rm fm}$, full angular ordering, red, solid; 2)
$L=0.5\,{\rm fm}$,
 partial angular ordering, green, dashed; 3) $L=5\,{\rm fm}$, partial angular ordering, blue, dotted.}
\label{paonl}
\end{figure}

In Fig.~\ref{paonl} we plot the distributions of final prehadrons
in the rapidity $y=\ln(E_0/E)$ for gluon jets with the initial
energy $E_0=100$ GeV and $L=$ 0.5, 1 and 5 fm.

From the distributions in Fig.~\ref{paonl} we see that the effect of color decoherence is indeed quite strong. Already at $L=$ 0.5 fm the
distribution is significantly different from the vacuum one and at larger $L$ practically saturates. Thus, even in the absence of energy loss in
terms of energy-momentum, the intrajet properties are significantly affected by the medium through color decoherence. The dominant effect of
decoherence seen in Fig.~\ref{paonl} is the softening of rapidity distribution accompanied by noticeable increase in the yield of hard final
particles and an overall relative flattening as compared to the in-vacuum case.

Two other results obtained in \cite{LN10} are:
\begin{itemize}
\item{Decoherence effects lead to a dramatic increase in the yield of prehadrons formed inside the hot zone.}
\item{Taking into account collisional energy losses of partons and prehadrons leads to significant depletion of the
multiplicity of final particles, noticeable overall energy loss and significant hardening of the rapidity distribution of final particles.}
\end{itemize}

The problem of medium-induced effects on quantum coherence is of significant importance for working out quantitative in-medium QCD. For recent
intriguing results see, e.g., \cite{MSK10}.

\begin{center}
{\bf Acknowledgements}
\end{center}

The work was supported by the RFBR grants 06-02-17051, 08=02-91000, 09-02-00741
and the RAS program "Physics at the LHC collider".

\end{document}